\documentclass[graybox]{svmult}

\usepackage{type1cm}      
 \usepackage{makeidx}         
\usepackage{graphicx}        
\usepackage{multicol}        
\usepackage[bottom]{footmisc}
\usepackage{newtxtext}   
\usepackage{newtxmath}       
\usepackage{color}

\newcommand{\be}{\begin{equation}}
\newcommand{\ee}{\end{equation}}
\newcommand{\bea}{\begin{eqnarray}}
\newcommand{\eea}{\end{eqnarray}}
\newcommand{\bean}{\begin{eqnarray*}}
\newcommand{\eean}{\end{eqnarray*}}
\newcommand{\one}{1\!\!{\rm I}}

\newcommand{\bra}{\langle}
\newcommand{\ket}{\rangle}

\newcommand{\order}{{\mathcal O}}

\newcommand{\N}{{\rm I\!N}}
\newcommand{\R}{{\rm I\!R}}

\newcommand{\bs}{\mbox{\protect\boldmath $s$}}

\newcommand{\bx}{\mbox{\protect\boldmath $x$}}
\newcommand{\by}{\mbox{\protect\boldmath $y$}}

\newcommand{\bA}{\mbox{\protect\boldmath $A$}}

\newcommand{\bK}{\mbox{\protect\boldmath $K$}}
\newcommand{\bL}{\mbox{\protect\boldmath $L$}}

\newcommand{\bQ}{\mbox{\protect\boldmath $Q$}}

\newcommand{\bDelta}{\mbox{\protect\boldmath $\Delta$}}

\newcommand{\bOmega}{\mbox{\protect\boldmath $\Omega$}}

\newcommand{\Internal}{\mathcal{E}}
\newcommand{\Partition}{\mathcal{Z}}
\newcommand{\Free}{\mathcal{F}}

\newcommand{\rmd}{{\rm d}}
\newcommand{\rme}{{\rm e}}
\newcommand{\rmi}{{\rm i}}
\newcommand{\notdelta}{\overline{\delta}}

\newcommand{\TE}{{\mathcal E}}

\makeindex           

\begin{document}

\title*{Dynamical replica analysis\\ of quantum annealing}
\author{ACC Coolen and T Nikoletopoulos}
\institute{ACCC and TN: Dept of Biophysics, Faculty of Science, Radboud University, 6525AJ Nijmegen, The Netherlands, and Saddle Point Science Ltd, United Kingdom. \email{a.coolen@science.ru.nl}
%\and 
%T Nikoletopoulos \at Dept of Biophysics Radboud University, 6525AJ Nijmegen, The Netherlands, and Saddle Point Science Ltd, United Kingdom, \email{t.nikoletopoulos@saddlepointscience.com}
}
\maketitle

\abstract{Quantum annealing aims to provide a faster method for finding the minima of complicated functions, compared to classical computing, so there is an increasing interest in the relaxation dynamics of quantum spin systems. Moreover, it is known that problems in quantum annealing caused by first order phase transitions can be reduced via appropriate temporal adjustment of  control parameters, aimed at steering the system away from local minima. To do this optimally, it would be helpful to predict the evolution of the system  at the level of macroscopic observables. Solving the dynamics of a quantum ensemble  is nontrivial, as it requires modelling not just the quantum spin system itself but also its interaction with the environment, with which it exchanges energy. An interesting alternative approach to the dynamics of quantum spin systems was proposed about a decade ago. It involves creating a stochastic proxy dynamics via the Suzuki-Trotter mapping of the quantum ensemble to a classical one (the quantum Monte Carlo method), and deriving from this new dynamics closed macroscopic equations for macroscopic observables, using the dynamical replica method.  In this chapter we give an introduction to this approach, focusing on the ideas and assumptions behind the derivations, and on its potential and limitations.  }

\section{Quantum ensembles and their dynamics}

We imagine an ensemble of $K$ independent quantum systems $|\psi^\alpha\ket$, labelled by $\alpha=1\ldots K$, all with the same Hamiltonian but distinct initial conditions. Making a measurement of an observable $A$ in this ensemble means picking randomly one of the $K$ systems, with equal probabilities, and measuring $A$ in the selected system. The average of the observable $A$ can then be written as $\bra A\ket= {\rm Tr}(\rho A)$, where $\rho$, the density matrix, is the Hermitian nonnegative definite operator 
$\rho= K^{-1}\sum_{\alpha=1}^K |\psi^\alpha\ket\bra \psi^\alpha  |$,
 with ${\rm Tr}(\rho) =1$. 
Since $\rho$ is Hermitian it has a complete basis of eigenstates $\{|k\ket\}$. Its eigenvalues $w_k$, which are nonnegative and normalised according to $\sum_k w_k=1$,  can be interpreted as probabilities. One can now write $\bra A\ket =\sum_n a_n \sum_k w_k |\bra k|n\ket|^2$.
Hence the probability to measure eigenvalue $a_n$ of observable $A$ in the ensemble is 
$P_n=\sum_k w_k |\bra k|n\ket|^2$, 
where $|\bra k|n\ket|^2$ is the probability to observe $a_n$ in eigenstate $k$ of the density matrix, and $w_k$ is the probability to find the ensemble in eigenstate $k$.

The evolution of the density matrix follows from that of the states $|\psi^\alpha\ket$, each governed by the Schr\"{o}dinger equation, giving
$\frac{\rmd}{\rmd t}\rho= (\rmi\hbar)^{-1}[H,\rho]$.
The solution is 
$\rho= \rme^{-\rmi Ht/\hbar} \rho_{t=0}~\rme^{\rmi Ht/\hbar}$. 
It follows in particular, using the eigenbasis $\{|E\ket\}$ of $H$, that 
\begin{eqnarray}
\bra H\ket &=& \sum_E \bra E| \rme^{-\rmi Ht/\hbar} \rho_{t=0}~\rme^{\rmi Ht/\hbar}H|E\ket
 ~=~\bra H\ket_{t=0}.
\label{eq:energy_over_time}
\end{eqnarray}
In equilibrium $[H,\rho]=0$, so the density matrix can be diagonalized simultaneously with $H$, i.e. 
$\rho=\sum_E f(E)|E\ket\bra E|$. 
The values of $f(E)$ define the type of equilibrium ensemble at hand. In the canonical ensemble we have $f(E)=\exp(-\beta E)/\Partition(\beta)$, so 
\begin{eqnarray}
\rho&=&\frac{1}{\Partition(\beta)}\sum_E \rme^{-\beta E}|E\ket\bra E|~=~\frac{1}{\Partition(\beta)}\rme^{-\beta H}.
\label{eq:canonical}
\end{eqnarray}
The quantum partition function $\Partition(\beta)$ follows from ${\rm Tr}(\rho)=1$:
$~\Partition(\beta)= {\rm Tr}(\rme^{-\beta H})$.
The free energy and the average internal energy are given by
$\Free=-\beta^{-1}\log\Partition(\beta)$ and $\Internal=-\frac{\partial}{\partial\beta}\log \Partition(\beta)$.
Expectation values  become
$\bra A\ket= \Partition(\beta)^{-1}{\rm Tr}(\rme^{-\beta H} A)$.
Note that if the systems of the ensemble evolve strictly according to the Schr\"{o}dinger equation, there cannot be generic evolution of $\rho$  towards the equilibrium form (\ref{eq:canonical}).  
For any initial density operator with $\bra H\ket_{t=0} \neq \Internal$ this is ruled out by  (\ref{eq:energy_over_time}).  
The state (\ref{eq:canonical}) describes the result of equilibration of the quantum systems in a heat bath, with which they can exchange energy, so a correct description of the dynamics would require a Hamiltonian that describes also the degrees of freedom of this heat bath. 

This is the first obstacle in the analysis of the dynamics of quantum ensembles: it is hard even to write down the correct microscopic  
dynamical laws. 
A similar situation occurs also in the classical setting. Without a heat bath we have a micro-canonical ensemble with conserved energy. Deriving the Gibbs-Boltzmann distribution from the joint dynamics of system and heat bath, requiring us to connect deterministic trajectories to invariant measures via ergodic theory and to subsequently derive the form of these measures,  has so far proven possible for only a handful of models. 

The approach followed in \cite{Nishimori}  was to circumvent ensembles altogether, and solve the Schr\"{o}dinger equation for small systems in which a decaying longitudinal field acts as quantum noise (which is indeed what happens in quantum annealing).  
In classical systems one often {\em defines} the pain away. One constructs an intuitively reasonable stochastic process that evolves towards the Gibbs-Boltzmann state, usually of the Markov Chain Monte Carlo (MCMC) form. This process is studied as a proxy for the dynamics of the original system.  
 The price paid is that one cannot be sure to what extent the stochastic dynamics is close to that of the original system. The MCMC equations are not even unique; many choices evolve to the Gibbs-Boltzmann state.  
 The same dynamics strategy can be applied to quantum systems if the latter can be mapped to classical ones. This is achieved by the Suzuki-Trotter formalism \cite{Suzuki}.
 
\section{Quantum Monte Carlo dynamics}

In applying quantum annealing to optimization problems formulated in terms of binary variables, one needs spin-$\frac{1}{2}$ particles \cite{Nishimori}. These are  labelled by $i=1\ldots N$, with  Pauli matrices $\{\sigma_{i}^x,\sigma_{i}^y,\sigma_i^z\}$. In the standard representation of $\sigma^z$-eigenstates: 
\begin{displaymath}
\sigma^x=\Big(\begin{array}{cc}0 & ~1\\ 1 & ~0\end{array}\Big),~~~~~~ \sigma^y=\Big(\begin{array}{cc}0 & -\rmi\\ \rmi & ~0\end{array}\Big),~~~~~~\sigma^z=\Big(\begin{array}{cc}1 & ~0\\ 0 & -1\end{array}\Big).
\end{displaymath}
In quantum annealing one chooses Hamiltonians of the form $H=H_0+H_1$, in which $H_0$ is obtained by replacing the classical spins $\sigma_i=\pm 1$ in an Ising Hamiltonian by the matrices $\sigma_i^z$, and with a second part $H_1$ that acts as a form of quantum noise\footnote{For simplicity we here choose $H_0$ to be quadratic in the spins, and the external field to be uniform, but this is not essential.  
}:
\begin{eqnarray}
H_0=-\sum_{i<j}J_{ij}\sigma_i^z \sigma_j^z-h\sum_i \sigma_i^z,~~~~~~~~H_1=-\Gamma \sum_{i}\sigma_i^x.
\label{eq:H0H1}
\end{eqnarray}
$H_0$ represents the quantity to be minimized in our optimization problem. The classical state achieving this minimum follows from the quantum ground state of the system upon sending the parameters $\Gamma$ and $\beta^{-1}$ adiabatically slowly to zero,  
and is hence obtained from the partition function 
$\Partition(\beta)= {\rm Tr}(\rme^{-\beta H_0-\beta H_1})$. For excellent reviews of the physics and the applications of the above types of quantum spin systems with transverse fields we refer to \cite{InoueReview,InoueBook}.

The Suzuki-Trotter procedure \cite{Suzuki} allows us to convert the above quantum problem into a classical one, using the operator identity
\begin{eqnarray}
\rme^{A+B}=\lim_{M\to\infty}\Big(\rme^{A/M}\rme^{B/M}\Big)^M.
\end{eqnarray}
From now on we assume that $A$ and $B$ are Hermitian operators, and we 
write the  basis of eigenstates of $A$ as $\{|n\ket\}$. We then obtain after some simple manipulations:
\begin{eqnarray}
{\rm Tr}(\rme^{A+B})&=&
\lim_{M\to\infty}\sum_{n_1 \ldots n_{M}}\rme^{\sum_{k=1}^M a_{n_k}/M}\!\!\prod_{k,~{\rm mod}(M)}\!\!
\bra n_k|\rme^{B/M}|n_{k+1}\ket.
\end{eqnarray}
Application to  $A=-\beta H_0$ and $B=-\beta H_1$, where the relevant basis is that of the joint eigenstates of all $\{\sigma_i^z\}$, i.e. $|s_1,\ldots,s_N\ket=|s_1\ket\otimes\ldots\otimes|s_N\ket$, with $s_i=\pm 1$ and 
$\sigma_i^z |s_1,\ldots,s_N\ket=s_i |s_1,\ldots,s_N\ket$, gives
$\Partition(\beta)=\lim_{M\to\infty}\Partition_M(\beta)$, where
\begin{eqnarray}
\Partition_M(\beta)&=& \!\!
 \sum_{\{s_{ik}=\pm 1\}}\!\!
\rme^{(\beta/M)\sum_{k=1}^M[ \sum_{i<j}J_{ij}s_{ik}s_{jk}+h\sum_i s_{ik}]}
\!\!
\prod_{k,~{\rm mod}(M)}\prod_{i=1}^N 
\bra s_{ik}|\rme^{(\beta\Gamma/M)\sigma_i^x}|s_{i, k+1}\ket
\nonumber
\\
&=& \rme^{
\frac{1}{2}NM\log[ \frac{1}{2}\sinh(2\beta\Gamma/M)]}
\nonumber
\\[0.5mm]&&\times\!\!
 \sum_{\{s_{ik}=\pm 1\}}
\rme^{(\beta/M)\sum_{k=1}^M [\sum_{i<j}J_{ij}s_{ik}s_{jk}+h\sum_is_{ik}]
+B\sum_{k,{\rm mod}(M)}\sum_{i} s_{ik}s_{i,k+1}}.~~~~
\end{eqnarray}
in which $B=-\frac{1}{2}\log  \tanh(\beta\Gamma/M)$. Thus the partition function of the $N$-spin quantum system is mapped (apart from a constant) onto the limit $M\to\infty$ of that of a classical Ising model with $NM$ spins $\bs=\{s_{ik}\}$, with Hamiltonian $H(\bs)$ and asymptotic free energy density $f= \lim_{N\to\infty}\lim_{M\to\infty}f_{N,M}$:
\begin{eqnarray}
H(\bs)&=&-\frac{1}{M}\sum_{k=1}^M \sum_{i<j}J_{ij}s_{ik}s_{jk}
-\frac{h}{M}\sum_{k=1}^M\sum_i s_{ik}
-\frac{B}{\beta}\sum_{k,{\rm mod}(M)}\sum_{i} s_{ik}s_{i,k+1},
\label{eq:Trotter_Hamiltonian}
\\
f_{N,M}&=& 
-\frac{M}{2\beta}
\log[ \frac{1}{2}\sinh(2\beta\Gamma/M)]
\nonumber
\\
&&\hspace*{0mm}
-\frac{1}{\beta N}\log \!\!
 \sum_{\{s_{ik}=\pm 1\}}\!\!
\rme^{\frac{\beta}{M}\sum_{k=1}^M [\sum_{i<j}J_{ij}s_{ik}s_{jk}+h\sum_i s_{ik}]
+B\sum_{k,{\rm mod}(M)}\sum_{i} s_{ik}s_{i,k+1}}.~~~~
\end{eqnarray}

The new system (\ref{eq:Trotter_Hamiltonian}), for $M\!\to\!\infty$ equivalent to the original quantum one, lends itself for constructing a stochastic dynamics.   
We first write the Suzuki-Trotter Hamiltonian in the standard form of $NM$ interacting Ising spins in an external field:
\begin{eqnarray}
&&
H(\bs)=-\frac{1}{2}\sum_{ik,j\ell}s_{ik} J_{ik,j\ell}s_{j\ell}-\theta\sum_{ik}s_{ik},
\label{eq:Trotter_H}
\\[-1mm]
&& J_{ik,j\ell}=\frac{1}{M}\delta_{k\ell}J_{ij}(1\!-\!\delta_{ij})+\frac{B}{\beta}\delta_{ij}(\delta_{k,\ell+1}\!+\!\delta_{\ell,k+1}),~~~~~~\theta=h/M.
\label{eq:general_H}
\end{eqnarray}
The conventional  Glauber dynamics for this classical system to evolve towards the equilibrium state with the above Hamiltonian is, after switching  to continuous time \cite{Bedeaux} and denoting with $p_t(\bs)$  the probability to find the system at time $t$ in state $\bs$:
\begin{eqnarray}
&&
\tau\frac{\rmd}{\rmd t}p_t(\bs)=\sum_{i=1}^N\sum_{k=1}^M \Big\{ p_t(F_{ik}\bs)w_{ik}(F_{ik}\bs)-p_t(\bs)w_{ik}(\bs)\Big\},
\label{eq:qMCMC1}
\\[-1mm]
&&
w_{ik}(\bs)=\frac{1}{2}[1-s_{ik}\tanh(\beta h_{ik}(\bs))],~~~~~~h_{ik}(\bs)=\sum_{j\ell}J_{ik,j\ell}s_{j\ell}+\theta.
\label{eq:qMCMC2}
\end{eqnarray}
It describes a process where at each step a site $i\!\in\!\{1,\ldots,N\}$ and a Trotter slice $k\!\in\!\{1,\ldots,M\}$ are picked at random, followed by an attempt to flip spin $s_{ik}$. 
The $w_{ik}(\bs)$ denote transition rates for $s_{ik}\to -s_{ik}$. $F_{ik}$ is an operator that flips spin $s_{ik}$ and leaves all others invariant. The parameter $\tau$ defines time units such that the average duration of a single spin update is $\tau/N$.  
Working out the local fields $h_{ik}(\bs)$ gives
\begin{eqnarray}
h_{ik}(\bs)=\frac{1}{M}\sum_{j\neq i}J_{ij} s_{jk}
 +\frac{B}{\beta}( s_{i, k+1}+s_{i, k-1})+h/M.
\end{eqnarray}
The process (\ref{eq:qMCMC1},\ref{eq:qMCMC2}), suitable for numerical simulation, defines the quantum Monte Carlo dynamics for the ensemble with Hamiltonian  (\ref{eq:H0H1}), provided we take $M\!\to\!\infty$. When applied to quantum annealing models, some authors have called it  `simulated quantum annealing'.  
Definition (\ref{eq:qMCMC1},\ref{eq:qMCMC2})  is, however, not unique. Many alternative stochastic processes evolve towards the same Gibbs-Boltzmann state (see e.g. \cite{Ohzeki}).

\section{Dynamical replica analysis}

The remaining challenge is to extract from (\ref{eq:qMCMC1},\ref{eq:qMCMC2})  formulae describing the evolution of relevant macroscopic quantities.  This was addressed in \cite{Inoue1,Inoue2,Bapst} and \cite{Arai} using the so-called dynamical replica method (DRT) \cite{DRT1,DRT2,DRT3}. In this paper we will deviate from the definitions in \cite{Inoue1,Inoue2,Bapst,Arai} and stay closer to the original DRT ideas.

The dynamics (\ref{eq:qMCMC1},\ref{eq:qMCMC2}) implies for expectation values $\bra G(\bs)\ket=\sum_{\bs}p_t(\bs)G(\bs)$:
\begin{eqnarray}
\tau
\frac{\rmd}{\rmd t}\bra G(\bs)\ket&=& \sum_{i=1}^N\sum_{k=1}^M  \sum_{\bs} p_t(\bs)w_{ik}(\bs)\Big[G(F_{ik}\bs)- G(\bs)\Big].
\end{eqnarray}
To study the joint dynamics of a set of $L$ observables $\bOmega(\bs)=(\Omega_1(\bs),\ldots,\Omega_L(\bs))$ we substitute 
$G(\bs)=\delta[\bOmega-\bOmega(\bs)]$. Now $
\bra G(\bs)\ket=P_t(\bOmega)$, and 
\begin{eqnarray}
\tau\frac{\rmd}{\rmd t} P_t(\bOmega)&=&\sum_{i=1}^N\sum_{k=1}^M  \sum_{\bs} p_t(\bs)w_{ik}(\bs)\Big[\delta[\bOmega-\bOmega(F_{ik}\bs)]- \delta[\bOmega-\bOmega(\bs)]\Big].
\label{eq:OmegaDens}
\end{eqnarray}
If the observables $\Omega_\mu(\bs)$ are $\order(1)$ and macroscopic in nature, their susceptibility to single spin flips $\Delta_{jk\mu}(\bs)=\Omega_{\mu}(F_{ik}\bs)-\Omega_\mu(\bs)$ will be small. We can then define $\bDelta_{jk}=(\Delta_{jk1}(\bs),\ldots,\Delta_{jkL}(\bs))\in\R^L$, and expand (\ref{eq:OmegaDens}) in a distributional sense, i.e.
\begin{eqnarray}
\tau\frac{\rmd}{\rmd t}\!\int\!\!\rmd\bOmega~P_t(\bOmega)G(\bOmega)&=&
  \int\!\!\rmd\bOmega~ G(\bOmega)\sum_{\ell\geq 1}\frac{(-1)^\ell}{\ell!}
 \frac{\partial^\ell }{\partial\Omega_{\mu_1}\ldots\partial\Omega_{\mu_\ell}} 
\nonumber
\\[-1mm]
&&\hspace*{-23mm} \times \Bigg\{
 \sum_{\mu_1=1}^{L}\ldots\sum_{\mu_\ell=1}^L 
 \sum_{i=1}^N\sum_{k=1}^M \Big\bra w_{ik}(\bs)\delta[\bOmega\!-\!\bOmega(\bs)]
\Delta_{ik\mu_1}(\bs)\ldots \Delta_{ik\mu_\ell}(\bs)\Big\ket
\Bigg\}.~~~~~~~
 \end{eqnarray}
 We thereby arrive at the following Kramers-Moyal expansion
 \begin{eqnarray}
\tau \frac{\rmd}{\rmd t}P_t(\bOmega)
 &=& 
 \sum_{\ell\geq 1}\frac{(-1)^\ell}{\ell!}\sum_{\mu_1=1}^{L}\ldots\sum_{\mu_\ell=1}^L \frac{\partial^\ell}{\partial\Omega_{\mu_1}\ldots\partial\Omega_{\mu_\ell}}\Big\{P_t(\bOmega)F^{(\ell)}_{\mu_1\ldots\mu_\ell}[\bOmega;t]\Big\},~~
 \label{eq:KM}
\end{eqnarray}
with\\[-12mm]
\begin{eqnarray}
F^{(\ell)}_{\mu_1\ldots\mu_\ell}[\bOmega;t]&=& \Bigg\bra \sum_{i=1}^N\sum_{k=1}^M w_{ik}(\bs)\Delta_{ik\mu_1}(\bs)\ldots \Delta_{ik\mu_\ell}(
\bs)\Bigg\ket_{\!\bOmega;t},
\\
\bra f(\bs)\ket_{\bOmega;t}&=& \frac{\sum_{\bs}p_t(\bs)\delta[\bOmega-\bOmega(\bs)]f(\bs)}{\sum_{\bs}p_t(\bs)\delta[\bOmega-\bOmega(\bs)]}.
\end{eqnarray}
Asymptotically, i.e. for $N,M\to\infty$,  only the first term of (\ref{eq:KM}) survives if 
\begin{eqnarray}
\lim_{N,M\to\infty}
 \sum_{\ell\geq 2}\frac{1}{\ell!}\sum_{\mu_1=1}^{L}\ldots\sum_{\mu_\ell=1}^L 
  \sum_{i=1}^N\sum_{k=1}^M \Big\bra |\Delta_{ik\mu_1}(\bs)\ldots \Delta_{ik\mu_\ell}(
\bs)|\Big\ket_{\bOmega;t}=0.
\end{eqnarray}
If all $\Delta_{ik\mu}(\bs)$ scale similarly, i.e. $\exists \tilde{\Delta}_{N,M}$ such that $\Delta_{ik\mu}(\bs)=\order(\tilde{\Delta}_{N,M})$ for $N,M\!\to\!\infty$, then (\ref{eq:KM}) retains only its first term if
 $\lim_{N,M\to\infty} L\tilde{\Delta}_{N,M}\sqrt{NM}=0$. In that case it  
   reduces to a Liouville equation, describing deterministic evolution of $\bOmega$:
\begin{eqnarray}
\tau\frac{\rmd}{\rmd t}\Omega_\mu&=&  \Big\bra \sum_{i=1}^N\sum_{k=1}^M w_{ik}(\bs)\Delta_{ik\mu}(\bs)\Big\ket_{\bOmega;t}.
\label{eq:macroscopic_flow}
\end{eqnarray}
 If $\lim_{N,M\to\infty} L\tilde{\Delta}_{N,M}\sqrt{NM}>0$, we can no longer ignore the fluctuations in our observables $\bOmega(\bs)$, placing limitations on our choice of observables.

Equation (\ref{eq:macroscopic_flow}) is closed if $\sum_{i=1}^N\sum_{k=1}^M w_{ik}(\bs)\Delta_{ik\mu}(\bs)$ is a function of $\bOmega(\bs)$ only (which would simply drop out). If this is not the case, we close (\ref{eq:macroscopic_flow}) using a maximum entropy argument:  we approximate $p_t(\bs)$ in (\ref{eq:macroscopic_flow}) by a form that assumes that all micro-states with the same value for $\bOmega(\bs)$ are equally likely.  Now  (\ref{eq:macroscopic_flow})  becomes
\begin{eqnarray}
\tau\frac{\rmd}{\rmd t}\Omega_\mu&=& \frac{\sum_{\bs}\delta[\bOmega-\bOmega(\bs)] \sum_{i=1}^N\sum_{k=1}^M w_{ik}(\bs^1)\Delta_{ik\mu}(\bs)}
{\sum_{\bs}\delta[\bOmega-\bOmega(\bs)]}.
\label{eq:closed_macroscopic_flow}
\end{eqnarray}
Within the replica formalism \cite{replicas1,replicas2},  this closed equation can also be written as
\begin{eqnarray}
\tau\frac{\rmd}{\rmd t}\Omega_\mu&=&
 \lim_{n\to 0}\sum_{\bs^1 \ldots\bs^n}\Bigg(\prod_{\alpha=1}^n \delta[\bOmega\!-\!\bOmega(\bs^\alpha)\Bigg)
\sum_{i=1}^N\sum_{k=1}^M w_{ik}(\bs^1)\Delta_{ik\mu}(\bs^1).
\label{eq:closed_macroscopic_flow_replicas}
\end{eqnarray}
The accuracy of (\ref{eq:closed_macroscopic_flow}) will depend on our choice for the observables $\Omega_\mu(\bs)$.  They should be $\order(1)$, obeying $\lim_{N,M\to\infty} L\tilde{\Delta}_{N,M}\sqrt{NM}=0$, and such that the probability equipartitioning assumption is as harmless as possible. Including $H(\bs)/N$ and $N^{-1}\log p_0(\bs)$ ensures that equipartitioning holds for $t\to 0$ and $t\to\infty$. 
If we have disorder in the couplings $\{J_{ij}\}$, and for $N\!\to\!\infty$ our observables are self-averaging with respect to its realization, we can average  over the disorder\footnote{Without disorder one does not need the replica formalism yet, and can work directly with (\ref{eq:closed_macroscopic_flow}).}. This gives
\begin{eqnarray}
\tau\frac{\rmd}{\rmd t}\Omega_\mu&=&
  \lim_{n\to 0}\sum_{\bs^1 \ldots\bs^n}
\overline{\Bigg(\prod_{\alpha=1}^n \delta[\bOmega\!-\!\bOmega(\bs^\alpha)\Bigg)
\sum_{i=1}^N\sum_{k=1}^M w_{ik}(\bs^1)\Delta_{ik\mu}(\bs^1)}.
\label{eq:closed_averaged_macroscopic_flow}
\end{eqnarray}

For the system (\ref{eq:Trotter_Hamiltonian}) and the typical initial conditions in quantum annealing, there are two natural and simple routes for choosing the observables in the DRT method\footnote{One can always add further observables, or split the present ones into distinct contributions. This generally improves the accuracy of the theory, provided  $\lim_{N,M\to\infty} L\tilde{\Delta}_{N,M}\sqrt{NM}=0$ still holds.}, all involving the normalised distinct energy contributions in  (\ref{eq:Trotter_Hamiltonian}):
\begin{itemize}
\item {\em Trotter slice dependent observables}
\\[2mm]
Here we choose, for $k=1\ldots M~({\rm mod}~M)$, 
\begin{eqnarray}
E_k(\bs)=-\frac{1}{N} \!\sum_{i<j}\! J_{ij}s_{ik}s_{jk},~~~~m_k(\bs)=\frac{1}{N}\!\sum_i\! s_{ik},~~~~\TE_k(\bs)=\frac{1}{N}\!\sum_{i}\! s_{ik}s_{i,k+1}.
~~~~~~\label{eq:slice-dep}
\end{eqnarray}
Now $L=3M$, and the susceptibilities of the observables to single spin flips are, using $\sum_j J_{ij}s_{jk}=\order(1)$ for all $k$ (required for an extensive Hamiltonian):
\begin{eqnarray}
\Delta_{ik}E_q(\bs)&=& 2N^{-1}\delta_{qk} s_{ik}\sum_{ j\neq i}J_{ij}s_{jk}=\order(N^{-1}),
\label{eq:delta_1a}
\\[-1.5mm]
\Delta_{ik}m_q(\bs)&=&-2N^{-1}\delta_{qk}s_{ik}=\order(N^{-1}),
\label{eq:delta_1b}\\[1.5mm]
\Delta_{ik}\TE_q(\bs)&=&-2N^{-1}s_{ik}(\delta_{qk}s_{i,k+1}+\delta_{k,q+1}s_{i,k-1})  =\order(N^{-1}).
\label{eq:delta_1c}
\end{eqnarray}
Hence $\tilde{\Delta}_{N,M}=N^{-1}$, so deterministic evolution requires that $M\ll N^{\frac{1}{3}}$ as $M,N\to\infty$. Hence, on choosing (\ref{eq:slice-dep}) we can no longer take $M\!\to\!\infty$ before $N\!\to\!\infty$, which would have been the correct order, and {must} rely on these limits commuting\footnote{The assumption that the order of the limits $N\!\to\!\infty$ and $M\!\to\!\infty$ can be changed is also made in equilibrium studies such as \cite{Nishimori2}, where steepest descent integration is used as $N\!\to\!\infty$ for fixed $M$.}.
\vspace*{2mm}

\item {\em Trotter slice independent observables}
\\[2mm]
These are simply averages over all Trotter slices of the previous set (\ref{eq:slice-dep}), i.e. 
\begin{eqnarray}
E(\bs)=\frac{1}{M}\!\sum_{k=1}^M\! E_k(\bs),~~~~~~m(\bs)=\frac{1}{M}\!\sum_{k=1}^M \! m_k(\bs),~~~~~~\TE(\bs)=\frac{1}{M}\!\sum_{k=1}^M \!\TE_k(\bs).
~~~~\label{eq:slice-indep}
\end{eqnarray}
Hence $L=3$, and the spin-flip susceptibilities come out as
\begin{eqnarray}
\Delta_{ik}E(\bs)&=& 2(NM)^{-1}s_{ik}\sum_{ j\neq i}J_{ij}s_{jk}=\order((NM)^{-1}),
\label{eq:delta_2a}\\[-1.5mm]
\Delta_{ik}m(\bs)&=&-2(NM)^{-1} s_{ik}=\order((NM)^{-1}),
\label{eq:delta_2b}\\[1.5mm]
\Delta_{ik}\TE(\bs)&=&-2(NM)^{-1}s_{ik}(s_{i,k+1}\!+\!s_{i,k-1}) =\order((NM)^{-1}).
\label{eq:delta_2c}
\end{eqnarray}
Now $\tilde{\Delta}_{N,M}=1/NM$.  
Deterministic evolution requires $\lim_{N,M\to \infty} (NM)^{-\frac{1}{2}}= 0$, which is always true. Here we can therefore take our two limits in any desired order without having to worry about fluctuations in our macroscopic observables. 
\end{itemize}

\section{Simple examples }

We illustrate the previous approach via application to simple models. We investigate the commutation of the limits $N\to\infty$ and $M\to\infty$, and the link between stationary states of the dynamical equations and the equilibrium theory.  We start with the simplest case of  non-interacting spins in a uniform $x$ field, followed by non-interacting spins in uniform $x$ and $z$ fields and ferromagnetically interacting quantum systems.

\subsection{Non-interacting quantum spins in a uniform $x$ field}

This is the simplest case of (\ref{eq:Trotter_Hamiltonian}), where $h=J_{ij}=0$ for all $(i,j)$. Although this specific model is physically trivial, it is still instructive since it already reveals many general features of the more general dynamical theory. The statics analysis gives
\begin{eqnarray}
\Partition_M(\beta)&=&\Big\{ \rme^{
\frac{1}{2}M\log[ \frac{1}{2}\sinh(2\beta\Gamma/M)]}
{\rm Tr}(\bK^M)
\Big\}^N,
\end{eqnarray}
with the $2\times 2$ transfer matrix  of the one-dimensional Ising chain:
\begin{eqnarray}
\bK=\left(\begin{array}{cc} \rme^B & \rme^{-B} \\
\rme^{-B} & \rme^B\end{array}
\right),~~~~~{\rm eigenvalues\!:}~~\lambda_+=2\cosh(B),~~\lambda_-=2\sinh(B).
\end{eqnarray}
After some rewriting and insertion of the definition of $B$ we obtain:
\begin{eqnarray}
\Partition_M(\beta)&=& \Big\{ \rme^{
\frac{1}{2}M\log[ \frac{1}{2}\sinh(2\beta\Gamma/M)]}
2^M [\cosh^M(B)+\sinh^M(B)]
\Big\}^N
\nonumber
\\
&=& [2\cosh(\beta\Gamma)]^N.
\end{eqnarray}
This gives the correct free energy density
$f_{N,M}=-\frac{1}{\beta}\log[2\cosh(\beta\Gamma)]$. 

Next we turn to the macroscopic dynamical equations (\ref{eq:macroscopic_flow}). Since $J_{ij}=0$, the order parameters $E_k(\bs)$ and $E(\bs)$ are always zero. The two dynamical routes give:
\begin{itemize}
\item {\em Trotter slice dependent observables}
\\[2mm]
The observables are $\{m_k(\bs),\TE_k(\bs)\}$, and we are forced to take $N\!\to\!\infty$ before $M\!\to\!\infty$. Using identities such as
$\tanh[B(s\!+\!s^\prime)]=\frac{1}{2}(s\!+\!s^\prime)\tanh(2B)$
we obtain:
\begin{eqnarray}
\tau\frac{\rmd}{\rmd t}m_k&=&-m_k+ \frac{1}{2}(m_{k+1}\!+\!m_{k-1})\tanh(2B),
\label{eq:zero1}
\\
\tau\frac{\rmd}{\rmd t}\TE_k&=&\tanh(2B)[1\!+\!\frac{1}{2}(C_{k}\!+\!C_{k+1})]-2\TE_k, 
\label{eq:zero2}
\end{eqnarray}
in which, using the equivalence of the $N$ sites $i$,  we have the 2-slice correlators:
\begin{eqnarray}
C_k&=& 
 \frac{\sum_{\bs}\Big[\prod_q \delta[m_q\!-\!m_q(\bs)]\delta[\TE_q\!-\!\TE_q(\bs)]\Big]
 s_{1, k-1}s_{1,k+1}}
{\sum_{\bs}\Big[\prod_q \delta[m_q\!-\!m_q(\bs)]\delta[\TE_q\!-\!\TE_q(\bs)]\Big]}.
\end{eqnarray}
One can compute these  for $N\to\infty$ with fixed $M$ via steepest descent integration:
\begin{eqnarray}
C_k&=& 
\frac{\sum_{s_1\ldots s_M}\rme^{\sum_{q}(x_q s_{q}+y_q s_{q}s_{q+1})}
 s_{k-1}s_{k+1}}{\sum_{s_1\ldots s_M}\rme^{\sum_{q}(x_q s_{q}+y_q s_{q}s_{q+1})}
},
\label{eq:Ck}
\end{eqnarray}
in which $\bx=(x_1,\ldots,x_M)$ and $\by=(y_1,\ldots,y_M)$ are to be solved from
\begin{eqnarray}
m_k=\frac{\partial \log Z}{\partial x_k},~~~~~\TE_k=\frac{\partial \log Z}{\partial y_k},~~~~~
Z(\bx,\by)=\!\!\!\sum_{s_1\ldots s_M}\!\!\!\rme^{\sum_{q}(x_q s_{q}+y_q s_{q}s_{q+1})}\!.~~~~
\end{eqnarray}

\item  {\em Trotter slice independent observables}
\\[2mm]
Here we only have $m(\bs)$ and $\TE(\bs)$, and working out (\ref{eq:macroscopic_flow}) gives
\begin{eqnarray}
\tau\frac{\rmd}{\rmd t}m= - m[1\!-\!\tanh(2B)],~~~~~~
\tau\frac{\rmd}{\rmd t}\TE=(1\!+\!C)\tanh(2B)-2\TE,
\label{eq:zero_0}
\end{eqnarray}
with\\[-8mm]
\begin{eqnarray}
C&=& 
 \frac{\sum_{\bs}\delta[m\!-\!m(\bs)]\delta[\TE\!-\!\TE(\bs)]
s_{1, 1}s_{1,3}}
{\sum_{\bs}\delta[m\!-\!m(\bs)]\delta[\TE\!-\!\TE(\bs)]}.
\end{eqnarray}
Calculating the 2-slice correlator $C$ using steepest descent results in
\begin{eqnarray}
&&\hspace*{15mm} C= 
\frac{\sum_{s_1\ldots s_M}\rme^{\frac{1}{M}\sum_{q}(x s_{q}+y s_{q}s_{q+1})}
 s_{1}s_{3}}{\sum_{s_1\ldots s_M}\rme^{\frac{1}{M}\sum_{q}(x s_{q}+y s_{q}s_{q+1})}
},
\label{eq:static_C}
\\[1mm]
&& \hspace*{-10mm} m=\frac{\partial \log Z}{\partial x},~~~~~~
\TE=\frac{\partial\log Z}{\partial y},~~~~~~Z(x,y)=\!\sum_{s_1\ldots s_M}\rme^{\frac{1}{M}\sum_{q}(x s_{q}+y s_{q}s_{q+1})}.
\label{eq:static_xy}
\end{eqnarray}
\end{itemize}
If at time zero the $m_k$ and $\TE_k$ in (\ref{eq:zero1},\ref{eq:zero2}) are independent of $k$, this will remain true at all times\footnote{In \cite{Inoue1,Inoue2,Arai} this is called the static approximation.} and the dynamics (\ref{eq:zero1},\ref{eq:zero2}) simplifies to (\ref{eq:zero_0}). 
Computing $C$ involves solving a one-dimensional Ising model with a constant external field, whereas computing the $C_k$ requires solving heterogeneous spin chain models in equilibrium for arbitrary coupling constants and fields.  This is the second reason, in addition to the issue with limits,  why working with Trotter slice independent observables is preferred.

For non-interacting spins with  $h\neq 0$ the analysis is similar. Here 
$f=\lim_{M\to\infty}f_{N,M}= -\beta^{-1}\log [2\cosh(\beta\sqrt{\Gamma^2\!+\!h^2})]$, with equilibrium magnetisation  
\begin{eqnarray}
m&=&-\partial f/\partial h=\tanh(\beta\sqrt{h^2\!+\!\Gamma^2})\frac{h}{\sqrt{h^2\!+\!\Gamma^2}},
\end{eqnarray}
and the Trotter slice independent observables are predicted to obey 
\begin{eqnarray}
\tau\frac{\rmd}{\rmd t}m&=& \frac{1}{2}(1\!-\!C)\tanh(\beta h/M)+
\frac{1}{2}Q_+(1\!+\!C)-m(1\!-\!Q_-),
\label{eq:twofields1}
\\
\tau\frac{\rmd}{\rmd t}\TE
&=& (1\!+\!C)Q_- \! +2Q_+ m
-2\TE, ~~
\label{eq:twofields2}
\end{eqnarray}
with $Q_\pm=\frac{1}{2}[\tanh(\beta h/M\!+\!2B)\!\pm\!\tanh(\beta h/M\!-\!2B)]$. Since $\lim_{h\to 0}Q_+=0$ and $\lim_{h\to 0}Q_-=\tanh(2B)$,  equations (\ref{eq:twofields1},\ref{eq:twofields2}) indeed revert back to (\ref{eq:zero_0}) for $h\to 0$.
We will inspect the fixed-points of (\ref{eq:twofields1},\ref{eq:twofields2}) after having also added spin interactions in the next section. Clearly, since $\lim_{M\to\infty}Q_+=\lim_{M\to\infty}(1\!-\!Q_-)=0$ the relaxation time of the system will diverge for $M\to\infty$, with closer inspection revealing that $\rmd m/\rmd t=\order(M^{-2})$. This makes physical sense: for large $M$, hence large $B$,  the Trotter slices increasingly prefer identical states, so state changes  (in a single slice) become rare as they require the mounting energetic costs of breaking the Trotter symmetry.

\subsection{Ferromagnetic $z$-interactions and uniform $x$ and $z$ fields}

Here we choose $h\neq 0$, $\Gamma\neq 0$, and $J_{ij}=J_0/N$ for all $i\neq j$, so the the quantum Hamiltonian  is $H=-(J_0/N)\sum_{i<j}\sigma_i^z \sigma_j^z-\sum_i(h\sigma_i^z+\Gamma\sigma_i^x)$. This is known as the Husimi-Temperley-Curie-Weiss model in a transverse field \cite{Chayes}. In the statics we find, after some simple manipulations and with the short-hand ${\rm D}z=(2\pi)^{-\frac{1}{2}}\rme^{-\frac{1}{2}z^2}\rmd z$:
\begin{eqnarray}
\Partition_M(\beta)&=&   \rme^{
\frac{1}{2}NM\log[ \frac{1}{2}\sinh(2\beta\Gamma/M)]-\frac{1}{2}\beta J_0}
\!
\int\!\Bigg[\prod_{k=1}^M{\rm D}z_k\Bigg]
\Bigg\{ {\rm Tr}\prod_{k=1}^M\bK\Big(z_k\sqrt{\frac{M}{\beta J_0 N}}\Big)\Bigg\}^N\!\!\!\!,~~~~~~~~~
\label{eq:ferro_general}
\\[-5mm]&&\nonumber
\end{eqnarray}
with the non-symmetric transfer  matrix
\begin{eqnarray}
\bK(x)= \left(\begin{array}{cc} \rme^{B+\beta h/M+\beta J_0 x/M} & \rme^{-B+\beta J_0 x/M}
\\
\rme^{-B-\beta J_0 x/M} & \rme^{B-\beta h/M-\beta J_0 x/M }
\end{array}
\right)
=\rme^{x(\beta J_0/M)\sigma^z} \bK(0).
\\[-5mm]&&\nonumber
\end{eqnarray}

We first turn to the statics of the model. 
It is not immediately clear whether or not the limits $N,M\!\to\!\infty$ in (\ref{eq:ferro_general}) commute.
Upon taking the limit $N\to\infty$ first, one obtains via steepest descent integration:
\begin{eqnarray}
\lim_{N\to\infty}f_{N,M}&=& 
-\frac{M}{2\beta}\log\Big[ \frac{1}{2}\sinh\Big(\frac{2\beta\Gamma}{M}\Big)\Big]
-\frac{1}{\beta}{\rm extr}_{\bx} \Bigg\{
\log {\rm Tr}\prod_{k=1}^M \bK(x_k)-\frac{\beta J_0}{2M}\bx^2\Bigg\}.
\nonumber\\[-2mm]&&
\end{eqnarray}
We find the derivatives of the quantity $\Psi(\bx)$ to be extremized, with $\notdelta_{ab}=1\!-\!\delta_{ab}$:
\begin{eqnarray}
\frac{\partial\Psi}{\partial x_q}&=& \frac{\beta J_0}{M}\Bigg\{
 \frac{{\rm Tr}\prod_{k=1}^M (\notdelta_{kq}\one \!+\!\delta_{kq}\sigma^z)\bK(x_k)} 
 {{\rm Tr}\prod_{k=1}^M  \bK(x_k)}-x_q\Bigg\},
\\
\frac{\partial^2\Psi}{\partial x_q \partial x_r}&=&
\Big(\frac{\beta J_0}{M}\Big)^2 \Bigg\{
 \frac{{\rm Tr}\prod_{k=1}^M (\notdelta_{kq}\one \!+\!\delta_{kq}\sigma^z)(\notdelta_{kr}\one \!+\!\delta_{kr}\sigma^z)\bK(x_k)} 
 {{\rm Tr}\prod_{k=1}^M  \bK(x_k)} \nonumber
 \\
 && \hspace*{-12mm} - 
 \frac{{\rm Tr}\prod_{k=1}^M\! (\notdelta_{kq}\one \!+\!\delta_{kq}\sigma^z)\bK(x_k)} 
 {{\rm Tr}\prod_{k=1}^M  \bK(x_k)}
  \frac{{\rm Tr}\prod_{k=1}^M\! (\notdelta_{kr}\one \!+\!\delta_{kr}\sigma^z)\bK(x_k)} 
 {{\rm Tr}\prod_{k=1}^M  \bK(x_k)} 
  \Bigg\}-\frac{\beta J_0}{M}\delta_{qr}.
  ~~~~~~~~
\end{eqnarray}
  In Trotter-symmetric solutions $x_k=m$ for all $k$, these derivatives simplify to
  \begin{eqnarray}
\frac{\partial\Psi}{\partial x_q}&=&\frac{\beta J_0}{M}\Bigg\{
 \frac{{\rm Tr}[ \sigma^z\bK^M\!(m)]} 
 {{\rm Tr} [\bK^M\!(m)]}-m\Bigg\},
\\
\frac{\partial^2\Psi}{\partial x_q \partial x_r}&=&
\Big(\frac{\beta J_0}{M}\Big)^2 \Bigg\{
 \frac{{\rm Tr}[\sigma^z\bK^{|q-r|}\!(m)\sigma^z \bK^{M-|q-r|}\!(m)] } 
 {{\rm Tr}[ \bK^M\!(m)]}
  - 
\Bigg(  \frac{{\rm Tr}[ \sigma^z\bK^M\!(m)]} 
 {{\rm Tr}[ \bK^M\!(m)]}
  \Bigg)^2\Bigg\}
  \nonumber
  \\[1mm]
  &&\hspace*{60mm} 
 -(\beta J_0/M) \delta_{qr}.
 \end{eqnarray}  
 and $m$ is the solution of \\[-8mm]
 \begin{eqnarray}
m=
 \frac{{\rm Tr}[ \sigma^z \bK^M\!(m)]} 
 {{\rm Tr}[ \bK^M\!(m)]}. \end{eqnarray} 
 Trotter symmetry-breaking bifurcations occur when ${\rm Det}[(\beta J_0/M) \bA\!-\!\one]=0$, where
 \begin{eqnarray}
 A_{qr}=
\frac{{\rm Tr}[\sigma^z \bK^{|q-r|}\!(m)\sigma^z \bK^{M-|q-r|}\!(m)] } 
 {{\rm Tr}[ \bK^M\!(m)]}-m^2.
\end{eqnarray}  
We introduce the symmetric  matrix $\bQ(m)=\rme^{-\frac{1}{2}m(\beta J_0/M)\sigma^z}\!\bK(m)\rme^{\frac{1}{2}m(\beta J_0/M)\sigma^z}\!$, with eigenvalues $\lambda_\pm (x)$ and orthogonal eigenbasis $|\pm\ket$. Now for any $\ell \in\N$ we have
 \begin{eqnarray}
 \bK^\ell(m) &=&  \rme^{\frac{1}{2}m(\beta J_0/M)\sigma^z}\Big(\lambda_+^\ell(m)|+\ket\bra +|
 \!+\!\lambda_-^\ell(m)|-\ket\bra -|\Big)\rme^{-\frac{1}{2}m(\beta J_0/M)\sigma^z},
 \end{eqnarray}
 and hence, with the short-hands $\sigma_{ab}^z=\bra a|\sigma^z|b\ket$ and $\phi=\lambda_-(m)/\lambda_+(m)\in(-1,1)$:
\begin{eqnarray}
 \frac{{\rm Tr}[ \sigma^z \bK^M\!(m)]} 
 {{\rm Tr}[ \bK^M\!(m)]}&=&  
\frac{\sigma^z_{++}\!+\!\sigma^z_{--} \phi^M} 
 {1+\phi^M}, 
 \\
A_{qr} &=&
\frac{\sigma^{z 2}_{++}\!
 +\!\big[\phi^{|q-r|} \!+\!\phi^{M-|q-r|}\big]|\sigma^z_{+-}|^2
\!+ \!\phi^M\sigma^{z2}_{--}} 
    {1+\phi^M}  - m^2.
\end{eqnarray} 
Since $\bA$ has a Toeplitz form, we know its eigenvalues:
\begin{eqnarray}
k=1\ldots M\!:~~~~a_{k}&=& \frac{|\sigma^z_{+-}|^2}  {1\!+\!\phi^M} \frac{(1\!-\!\phi^M)(1\!-\!\phi^2)}{1\!+\!\phi^2\!-\!2\phi\cos(2\pi (k\!-\!1)/M)}.
\end{eqnarray}
Finally we need to diagonalize $\bQ(m)$ for large $M$. This gives:
\begin{eqnarray}
\bQ(m)&=&
\left(\!\begin{array}{cc} 
\rme^{B+\beta(h+J_0m)/M} & \rme^{-B}\\
\rme^{-B} & \rme^{B-\beta(h+J_0m)/M} 
\end{array}\!\right)
\label{eq:Qmatrix}
\\
\lambda_\pm(m)&=&\rme^{B\pm \frac{\beta}{M}\sqrt{(h+J_0m)^2+\Gamma^2}
+\order(M^{-2})},
\\
\lim_{M\to\infty}|\pm\ket &=& \frac{1}{C_\pm(m)}\Big(\Gamma,-(h\!+\!J_0m)
\pm\sqrt{(h\!+\!J_0m)^2\!+\!\Gamma^2}\Big),
\\
C_{\pm}(m)&=& \sqrt{2}\Big[
(h\!+\!J_0m)^2\!+\!\Gamma^2\mp (h\!+\!J_0m)\sqrt{(h\!+\!J_0m)^2\!+\!\Gamma^2}\Big]^{\frac{1}{2}}.
\end{eqnarray}
It follows that\\[-8mm]
\begin{eqnarray}
\phi=\rme^{- \frac{2\beta}{M}\sqrt{(h+J_0m)^2+\Gamma^2}
+\order(M^{-2})}.
\end{eqnarray}
Hence $\lim_{M\to\infty}\phi=1$, $\lim_{M\to\infty}\phi^M=\exp[-2\beta\sqrt{(h\!+\!J_0m)^2\!+\!\Gamma^2}]$, $\lim_{M\to\infty}\sigma_{++}^z=-\lim_{M\to\infty}\sigma_{--}^z\!=\!(h\!+\!J_0m)/\sqrt{(h\!+\!J_0m)^2\!\!+\!\Gamma^2}$,  and 
$\lim_{M\to\infty}\sigma_{+-}^z\!=\!\Gamma/\sqrt{(h\!+\!J_0m)^2\!\!+\!\Gamma^2}$. The equation for the magnetization $m$ and the eigenvalues of $\bA$ thereby become 
\begin{eqnarray}
m&=& \frac{(h\!+\!J_0m)\tanh[\beta\sqrt{(h\!+\!J_0m)^2\!+\!\Gamma^2}]}{\sqrt{(h\!+\!J_0m)^2\!+\!\Gamma^2}},
\label{eq:m_ferro}
 \\
 a_{k}&=& \frac{\Gamma^2\tanh[\beta\sqrt{(h\!+\!J_0m)^2\!+\!\Gamma^2}]}{ (h\!+\!J_0m)^2\!+\!\Gamma^2}
 \Big[
1\!+\! 2\lim_{M\to\infty}\frac{1\!-\!\cos(2\pi (k\!-\!1)/M)}{1\!-\!\phi^2}\Big]^{-1}.
 \end{eqnarray}
 All $a_k$ are bounded for large $M$, so the condition $\beta J_0 a_k/M=1$ for bifurcations away from the Trotter-symmetric state are never met, indicating that the state described by (\ref{eq:m_ferro}) is the physical one. The free energy density $f=\lim_{M\to\infty}\lim_{N\to\infty}f_{N,M}$ is
 \begin{eqnarray}
f&=& \frac{1}{2}J_0 m^2
-\lim_{M\to\infty}\Bigg\{
\frac{M}{2\beta}\log\Big[ \frac{1}{2}\sinh\Big(\frac{2\beta\Gamma}{M}\Big)\Big]
+\frac{1}{\beta} 
\log
\Big(\lambda_+^M(m)
 \!+\!\lambda_-^M(m)\Big)
 \Bigg\}
\nonumber\\
&=&  \frac{1}{2}J_0 m^2-\frac{1}{\beta} 
\log
\Big[2\cosh\Big(\beta\sqrt{(h\!+\!J_0m)^2\!+\!\Gamma^2}\Big)\Big].
\label{eq:f_ferro}
 \end{eqnarray}
 Extremizing expression (\ref{eq:f_ferro}) over $m$ reproduces (\ref{eq:m_ferro}).

We return to (\ref{eq:ferro_general}), and now seek to take the Trotter limit $M\!\to\!\infty$ first. The complexities are all in the evaluation for large $M$ of the quantity
\begin{eqnarray}
Z_M&=& 
\int\!\Bigg[\prod_{k=1}^M{\rm D}z_k\Bigg]
\Bigg\{ {\rm Tr}\Bigg[\prod_{k=1}^M\rme^{z_k\sqrt{\frac{\beta J_0}{MN}}\sigma^z} \left(\begin{array}{cc} \rme^{B+\beta h/M} & \rme^{-B}\\
\rme^{-B} & \rme^{B-\beta h/M }
\end{array}
\right)\Bigg]
\Bigg\}^N.
\end{eqnarray}
This could be analysed using random field Ising chain techniques \cite{RFIC}. Alternatively we can use the fact that in summations of the form $\sum_k z_k$, each $z_k$ effectively scales as $\order(M^{-\frac{1}{2}})$, enabling us to use $\rme^{-B}=\sqrt{\tanh(\beta\Gamma/M)}$ and a modified version of the Trotter identity, viz. $\prod_{k\leq M}\big( \rme^{u_k/M}\rme^{v/M}\big)=\rme^{M^{-1}\sum_{k\leq M}u_k+v}$, to derive
\begin{eqnarray}
Z_M&=& 
\rme^{NMB}
\int\!\Bigg[\prod_{k=1}^M{\rm D}z_k\Bigg]
\Bigg\{ {\rm Tr}\Bigg[\prod_{k=1}^M\rme^{z_k\sqrt{\frac{\beta J_0}{MN}}\sigma^z} 
\Big(\one\!+\!\frac{\beta}{M}(h\sigma^z\!\!+\!\Gamma\sigma^x)\!+\!\order(\frac{1}{M^{2}})\Big)
\Bigg]
\Bigg\}^N
\nonumber
\\&=&\sqrt{\beta J_0 N}\rme^{NMB}
\int\!\!\frac{\rmd m}{\sqrt{2\pi}}\rme^{-\frac{1}{2}\beta J_0 Nm^2}~
\Big\{ {\rm Tr}~\rme^{ \beta (h+J_0 m)\sigma^z+\beta\Gamma\sigma^x\!+\order(M^{-1})}
\Big\}^N.
\end{eqnarray}
The free energy density $f=\lim_{N\to\infty}\lim_{M\to\infty}f_{N,M}$ then becomes
\begin{eqnarray}
f&=&
-\frac{1}{\beta }{\rm extr}_m  \Big\{
\log\big(\rme^{\beta\mu_+(m)}+\rme^{\beta\mu_-(m)}\big)
 -\frac{1}{2}\beta J_0 m^2\Big\},
 \end{eqnarray}
 in which $\mu_\pm(m)$ are the eigenvalues of the matrix $\bL(m)=(h\!+\!J_0m)\sigma^z\!+\!\Gamma\sigma^x$:
 \begin{eqnarray}
 \bL(m)=\left(\!\begin{array}{cc} h\!+\!J_0 m & \Gamma\\ \Gamma & -(h\!+\!J_0m)\end{array}\!\right),~~~~~~\mu_\pm(m)=\pm\sqrt{(h\!+\!J_0m)^2\!+\!\Gamma^2}.
 \end{eqnarray}
 We now recover (\ref{eq:m_ferro},\ref{eq:f_ferro}), so the limits $N\to\infty$ and $M\to\infty$ can be interchanged: 
 \begin{eqnarray}
 f&=&{\rm extr}_m  \Bigg\{\frac{1}{2} J_0 m^2
 -\frac{1}{\beta }\log\Big[2\cosh\Big(\beta\sqrt{(h\!+\!J_0m)^2\!+\!\Gamma^2}\Big)\Big]
 \Bigg\}.
 \end{eqnarray} 

We next turn to the DRT dynamics. 
The energy and the usual initial conditions can once more be expressed in terms of $\{m_k,\TE_k\}$ (slice-dependent observables) or $(m,\TE)$ (slice-independent ones). We define the short-hands
$Q_{\pm}(m)\!=\!\frac{1}{2}\tanh(\beta(J_0m\!+\!h)/M\!+\!2B)\!\pm\!\frac{1}{2}\tanh(\beta(J_0m\!+\!h)/M\!-\!2B)\in(-1,1)$. Upon inserting (\ref{eq:delta_1b},\ref{eq:delta_1c}) and (\ref{eq:delta_2b},\ref{eq:delta_2c}) into (\ref{eq:macroscopic_flow}), with the fields $h_{ik}(\bs)=M^{-1}[h\!+\! J_0m_k(\bs)]+(B/\beta)(s_{i,k+1}\!+\!s_{i,k-1})+\order(N^{-1})$, and using expressions such as $\tanh[a+b(s\!+\!s^\prime)]=\frac{1}{4}(1\!+\!s)(1\!+\!s^\prime)\tanh(a\!+\!2b)
+\frac{1}{4}(1\!-\!s)(1\!-\!s^\prime)\tanh(a\!-\!2b)+\frac{1}{2}(1\!-\!ss^\prime)\tanh(a)$, one finds the following descriptions:

\begin{itemize}
\item {\em Trotter slice dependent observables}
\\[2mm]
Our observables are $m_q(\bs)\!=\!N^{-1}\sum_i s_{i,q}$ and $\TE_q(\bs)\!=\!N^{-1}\sum_i s_{i,q}s_{i,q+1}$, for $q=1\ldots M$, and  
 we must take the limit $N\to\infty$ before $M\to\infty$. We note that 
\begin{eqnarray}
\tanh(\beta h_{ik}(\bs))&=& 
\frac{1}{2}(1\!+\!s_{i,k+1} s_{i,k-1})Q_+(m_k(\bs))+\frac{1}{2}(s_{i,k+1}\!+\!s_{i,k-1})Q_-(m_k(\bs))
\nonumber
\\&&+\frac{1}{2}(1\!-\!s_{i,k+1}s_{i,k-1})\tanh(\beta(h\!+\!J_0 m_k(\bs))/M),
\end{eqnarray}
so with the correlators $C_k$ in (\ref{eq:Ck}) the dynamical laws take the form  
 \begin{eqnarray}
\tau\frac{\rmd}{\rmd t}m_q&=& 
\frac{1}{2}(1\!+\!C_q)Q_+(m_q)+\frac{1}{2}(m_{q+1}\!+\!m_{q-1})Q_-(m_q)
-m_q
\nonumber
\\
&&+\frac{1}{2}(1\!-\!C_q)\tanh(\frac{\beta}{M}(h\!+\! J_0m_q)),
\\
\tau\frac{\rmd}{\rmd t}\TE_q&=&  
\frac{1}{2}(m_{q+1}\!+\!m_{q-1})Q_+(m_q)
+\frac{1}{2}(1\!+\!C_q)Q_-(m_q)
\nonumber
\\&&
+\frac{1}{2}(m_{q}\!+\!m_{q+2})Q_+(m_{q+1})
+\frac{1}{2}(1\!+\!C_{q+1})Q_-(m_{q+1})
\nonumber
\\&&
+\frac{1}{2}(m_{q+1}\!-\!m_{q-1})\tanh(\frac{\beta}{M}(h\!+\! J_0m_q))
\nonumber
\\&&
+\frac{1}{2}(m_{q}\!-\!m_{q+2})\tanh(\frac{\beta}{M}(h\!+\! J_0m_{q+1}))
-2\TE_q.
\end{eqnarray}
 For slice-independent initial conditions, where $m_k=m$ and $\TE_k=\TE$, this becomes
  \begin{eqnarray}
\tau\frac{\rmd}{\rmd t}m&=& 
\frac{1}{2}(1\!+\!C)Q_+(m)\!+\!mQ_-(m)
\!-\!m\!+\!\frac{1}{2}(1\!-\!C)\tanh(\frac{\beta}{M}(h\!+\! J_0m)),~~
\label{eq:ferroflow1}
\\
\tau\frac{\rmd}{\rmd t}\TE&=&  
2mQ_+(m)+(1\!+\!C)Q_-(m)
-2\TE,
\label{eq:ferroflow2}
\end{eqnarray} 
 with the correlator $C$ in (\ref{eq:static_C}). 
 \\[2mm]

\item {\em Trotter slice independent observables}
\\[2mm]
For the choice $(m,\TE)$ there is no constraint on the order of limits, but now the quantities $m_k(\bs)$ appearing inside $\tanh(\beta h_{ik}(\bs))$ can no longer be replaced by deterministic macroscopic observables, but must now be calculated. Using Trotter slice permutation symmetry wherever possible, one finds
\begin{eqnarray}
\tau\frac{\rmd}{\rmd t}m&=&  \frac{1}{2M}\sum_{k=1}^M\Big\bra [1\!+\!C _k(\bs)]Q_+(m_k(\bs))+[m_{k+1}(\bs)\!+\!m_{k-1}(\bs)]Q_-(m_k(\bs))
\nonumber
\\&&
\hspace*{0mm}+[1\!-\!C_k(\bs)]\tanh(\beta(h\!+\!J_0 m_k(\bs))/M)\Big\ket_{\!m,\TE} - m,
\end{eqnarray}

\begin{eqnarray}
\tau\frac{\rmd}{\rmd t}\TE&=& \frac{1}{M} \sum_{k=1}^M \Big\bra [m_{k+1}(\bs)\!+\!m_{k-1}(\bs)]Q_+(m_k(\bs))\Big\ket_{\!m,\TE}
\nonumber
\\&&
+
\frac{1}{M}\sum_{k=1}^M \Big\bra [1\!+\!C_k(\bs)]Q_-(m_k(\bs))\Big\ket_{\!m,\TE}
-2\TE,
\end{eqnarray}
with $C_k(\bs)=N^{-1}\sum_i s_{i,k+1}s_{i,k-1}$.  
For large $M$ and $N$, and in view of the interchangeability of the limits $M\to\infty$ and $N\to\infty$ in the equilibrium calculation, we may anticipate (and can indeed show) that we can neglect the fluctuations in the values of the $\{m_k(\bs)$\} and simply replace $m_k(\bs)\to m(s)+{\it o}(1)$ in the right-hand sides of  above equations, upon which these simplify to (\ref{eq:ferroflow1},\ref{eq:ferroflow2}). 
\end{itemize}

 \section{Link between statics and dynamics}
 
 Here we show that for $M\to\infty$ the stationary state of the  (\ref{eq:ferroflow1},\ref{eq:ferroflow2}) reproduces the equilibrium result 
 (\ref{eq:m_ferro}), as  it should. The fixed-point equations of (\ref{eq:ferroflow1},\ref{eq:ferroflow2}) are 
 \begin{eqnarray}
m&=& 
\frac{1}{2}(1\!+\!C)Q_+(m)+mQ_-(m)
+\frac{1}{2}(1\!-\!C)\tanh(\frac{\beta}{M}(h\!+\! J_0m)),
\label{eq:fp_m}
\\
\TE&=&  
mQ_+(m)+\frac{1}{2}(1\!+\!C)Q_-(m),
\label{eq:fp_E}
\end{eqnarray} 
with the correlator $C=C(m,\TE)\in(-1,1)$ to be solved from
\begin{eqnarray}
&&\hspace*{15mm} C= 
\frac{\sum_{s_1\ldots s_M}\rme^{\sum_{k=1}^M(x s_{k}+y s_{k}s_{k+1})}
 s_{1}s_{3}}{\sum_{s_1\ldots s_M}\rme^{\sum_{k=1}^M(x s_{k}+y s_{k}s_{k+1})}
},
\\[1mm]
&& \hspace*{-10mm} 
m=\frac{1}{M}\frac{\partial \log Z}{\partial x},~~~~~~
\TE=\frac{1}{M}\frac{\partial\log Z}{\partial y},~~~~~~Z(x,y)=\!\sum_{s_1\ldots s_M}\rme^{\sum_{k=1}^M(x s_{k}+y s_{k}s_{k+1})}.
\label{eq:xy}
\end{eqnarray}
We compute $Z(x,y)$ via the transfer matrix $\bK(x,y)$ with elements $K_{ss^\prime}=\rme^{\frac{1}{2}x(s+s^\prime)+yss^\prime}$. This gives $Z(x,y)=\lambda^M_+(x,y)\!+\!\lambda^M_-(x,y)$, where $\lambda_{\pm}(.)$ are the eigenvalues of $\bK(.)$,  
\begin{eqnarray}
\lambda_{\pm}(x,y)&=& \rme^{y}\Big(\cosh(x)\pm \sqrt{\sinh^2(x)\!+\!\rme^{-4y}}\Big).
\end{eqnarray}
For the equilbrium values of $(m,\TE)$, equations (\ref{eq:xy}) are solved by 
\begin{eqnarray}
x\!=\!\beta(h\!+\!J_0m)/M,~~~~~~y\!=\!B\!=\!-\frac{1}{2}\log\tanh(\frac{\beta\Gamma}{M}),~~~~{\rm so}~~\rme^{-4y}=\tanh^2(\frac{\beta\Gamma}{M}).
~~~~
\end{eqnarray}
This claim is confirmed by substituting these as ans\"{a}tze into the expressions given in the appendix. 
The key ingredient $\phi=\lambda_-/\lambda_+$ of our formulae then becomes
\begin{eqnarray}
\log\phi &=& 
-\frac{2\beta}{M}\sqrt{(h\!+\!J_0m)^2\!+\!\Gamma^2}+\order(M^{-3}).
\end{eqnarray}
Hence for $M\to\infty$ the formulae for $m$ and $\TE$ in (\ref{eq:xy}) become
\begin{eqnarray}
m=
 \frac{(h\!+\!J_0m)\tanh[\beta\sqrt{(h\!+\!J_0m)^2\!+\!\Gamma^2}]}{\sqrt{(h\!+\!J_0m)^2\!+\!\Gamma^2}},~~~~~~
\TE=1.
\label{eq:thermodynamic_state}
\end{eqnarray}
in which we recognize (\ref{eq:m_ferro}). 
For large $M$ one finds $Q_+(m)\!=\!\order(M^{-3})$ and $Q_-(m)\!=\!1\!-\!2(\beta\Gamma/M)^2\!+\!\order(M^{-3})$, so expansion of the fixed-point equations gives
\begin{eqnarray}
m&=& 
M(1\!-\!C)\frac{h\!+\! J_0m}{4\beta\Gamma^2}+\order(M^{-1}),
\label{eq:expanded1}
\\
\TE&=&  
\frac{1}{2}(1\!+\!C)[1-2(\beta\Gamma/M)^2]+\order(M^{-3}).
\label{eq:expanded2}
\end{eqnarray} 
The first equation implies that $C=1\!-\!\tilde{C}/M$ for $M\to\infty$, with $\tilde{C}=\order(1)$. In turn, this gives $\TE=1\!-\!\frac{\tilde{C}}{2M} \!+\!\order(M^{-2})$. What is left in our proof  is to show that $m$ obeys  
 \begin{eqnarray}
m&=& 
\frac{h\!+\! J_0m}{4\beta\Gamma^2}\lim_{M\to\infty} M(1\!-\!C).
\label{eq:condition}
\end{eqnarray}
We hence compute the correlator $C$ to order $M^{-1}$, using the identities in the appendix:
\begin{eqnarray}
C &=&
\bra +|\sigma^z |+\ket^2
+\frac{\cosh[(\frac{1}{2}M\!-\!2)\log\phi]}{\cosh[\frac{1}{2}M\log \phi]}\Big(1-\bra +|\sigma^z |+\ket^2\Big)
\nonumber
\\&=& \frac{(h\!+\!J_0m)^2}{(h\!+\!J_0m)^2\!+\!\Gamma^2}
+\frac{\cosh[\beta(1\!-\!4/M)\sqrt{(h\!+\!J_0m)^2\!+\!\Gamma^2}]}{\cosh[\beta\sqrt{(h\!+\!J_0m)^2\!+\!\Gamma^2}]}
\frac{\Gamma^2}{(h\!+\!J_0m)^2\!+\!\Gamma^2}\!+\!\order(\frac{1}{M^2})
\nonumber
\\&=& 1\!-\!\frac{1}{M}\tanh\Big[\beta\sqrt{(h\!+\!J_0m)^2\!+\!\Gamma^2}\Big]
\frac{4\beta\Gamma^2}{\sqrt{(h\!+\!J_0m)^2\!+\!\Gamma^2}}\!+\!\order(\frac{1}{M^2}).
\end{eqnarray}
We can now read off the value of $\tilde{C}$, and 
the condition (\ref{eq:condition}) is found to reduce to (\ref{eq:m_ferro}), so that it is indeed satisfied. This completes the demonstration that  for large $M$ the macroscopic equations 
(\ref{eq:ferroflow1},\ref{eq:ferroflow2})  indeed have the equilibrium state as their fixed-point. 

\section{Evolution on adiabatically separated timescales}

We return to the dynamical laws (\ref{eq:ferroflow1},\ref{eq:ferroflow2}). As noted earlier, these exhibit for large $M$ a divergent relaxation time for the magnetization, suggesting  that the dynamics will have distinct phases. The first phase is studied by choosing $\tau=\order(1)$. Using
\begin{eqnarray}
Q_+(m)=
\frac{4\beta^3\Gamma^2(J_0m\!+\!h)}{M^3}+\order(M^{-4}),~~~~~~
Q_-(m)
=
1\!-\!\frac{2\beta^2\Gamma^2}{M^2}+\order(M^{-4}),~~~~~
\end{eqnarray}
we here find that 
 \begin{eqnarray}
 m=m_0+\order(M^{-1}),
 ~~~~~~
\tau\frac{\rmd}{\rmd t}\TE=
1\!+\!C(m_0,\TE)
-2\TE+\order(M^{-1}).
\end{eqnarray} 
So on these timescales the magnetization will not change, whereas the Trotter energy will evolve to the solution of the fixed-point equation $\TE=\frac{1}{2}+\frac{1}{2}C(m_0,\TE)$, in which $C(m_0,\TE)$ is according to the appendix to be solved from 
the following equations:
\begin{eqnarray}
m
&=&-  \frac{\sinh(x)\tanh[\frac{1}{2}M\log\phi]}{\sqrt{\sinh^2(x)\!+\!\rme^{-4y}}},
\label{eq:dyn_closure_m}
\\
\TE
&=&
\frac{\sinh^2(x)}{\sinh^2(x)\!+\!\rme^{-4y}}
\!+\!\frac{\cosh[(\frac{1}{2}M\!\!-\!1)\log\phi]}{\cosh[\frac{1}{2}M\log \phi]}
\frac{\rme^{-4y}}{\sinh^2(x)\!+\!\rme^{-4y}},
\label{eq:dyn_closure_E}\\
C
&=&\frac{\sinh^2(x)}{\sinh^2(x)\!+\!\rme^{-4y}}\!+\!\frac{\cosh[(\frac{1}{2}M\!\!-\!2)\log\phi]}{\cosh[\frac{1}{2}M\log \phi]}
\frac{\rme^{-4y}}{\sinh^2(x)\!+\!\rme^{-4y}},
\label{eq:dyn_closure_C}
\end{eqnarray}
with $\phi=[\cosh(x)\!-\!\sqrt{\sinh^2(x)\!+\!\rme^{-4y}}]/[\cosh(x)+ \sqrt{\sinh^2(x)\!+\!\rme^{-4y}}]$. Inspection of these equations reveals that 
 the correct scaling with $M$ requires $(x,\rme^{-2y})\!=\!(u,v)/M$, with $u,v\!=\!\order(1)$. Now $\frac{1}{2}M\log\phi \!=\! -\sqrt{u^2\!+\!v^2 }\!+\!\order(M^{-2})$, $\TE=1\!-\!\tilde{\TE}/M\!+\!\order(M^{-2})$, and $C=1\!-\!2\tilde{\TE}/M\!+\!\order(M^{-2})$, in which $(u,v)$ are solved from
\begin{eqnarray}
m_0=\frac{u\tanh(\sqrt{u^2\!+\!v^2 })}{\sqrt{u^2\!+\!v^2}},~~~~~~
\tilde{\TE}=\frac{2v^2\tanh(\sqrt{u^2\!+\!v^2 })}{\sqrt{u^2\!+\!v^2}}.
\end{eqnarray}
To order $\order(M^{-1})$ the fixed-point equation for $\TE$ is  now solved, but to compute $\tilde{\TE}$ one needs higher orders in $M^{-1}$. Once  $\TE=1\!-\!\tilde{\TE}/M\!+\!\order(N^{-2})$ and $C(m,\TE)=1\!-\!2\tilde{\TE}/M\!+\!\order(M^{-2})$,  we find $\rmd m/\rmd t=\order(M^{-2})$ and $\rmd\TE/\rmd t=\order(M^{-2})$, so nothing evolves further macroscopically on these finite timescales.

To probe the macroscopic evolution of the system on larger timescales we need $\tau=\order(M^{-2})$, so on unit timescales  $\order(M^3N)$ spin flips in the Trotter system are attempted\footnote{This reflects the high energy cost of breaking Trotter symmetry to induce magnetization changes.}. With the choice $\tau=M^{-2}$, and upon defining $M(1\!-\!\TE)=\tilde{\TE}$ and $M(1\!-\!C)=\tilde{C}$,  the macroscopic laws (\ref{eq:ferroflow1},\ref{eq:ferroflow2}) become
 \begin{eqnarray}
\frac{\rmd}{\rmd t}m&=& 
\frac{1}{2}\tilde{C}\beta(h\!+\! J_0m)-2m\beta^2\Gamma^2
+\order(\frac{1}{M}),
\label{eq:slow_m}
\\
\frac{\rmd}{\rmd t}\tilde{\TE}&=&  4M\beta^2\Gamma^2
-M^2(2\tilde{\TE}\!-\!\tilde{C})-8\beta^3\Gamma^2m(J_0m\!+\!h)-2\beta^2\Gamma^2\tilde{C}
+\order(\frac{1}{M}).~~~~~~~
\label{eq:slow_E}\end{eqnarray} 
The quantity $\tilde{C}=\tilde{C}(m,\tilde{\TE})$ is to be solved together with $(x,y)$ from (\ref{eq:dyn_closure_m},\ref{eq:dyn_closure_E},\ref{eq:dyn_closure_C}). The relevant scaling is still $(x,\rme^{-2y})=(u,v)/M$, with $u,v=\order(1)$, but according to (\ref{eq:slow_E}) we now need more than just the leading order in $M^{-1}$. Using 
\begin{eqnarray}
\log\phi&=&  -\frac{2\sqrt{u^2\!+\!v^2}}{M}+\order(M^{-3}),
\end{eqnarray}
the equations for $\TE$ and $C$ take the form $\TE=\Xi_1(u,v)$ and $C=\Xi_2(u,v)$, where
\begin{eqnarray}
\Xi_\ell(u,v)&=&
\Big[{\sinh^2(\frac{u}{M})\!+\!\frac{v^2}{M^2}}\Big]^{-1}\Big[
\sinh^2(\frac{u}{M})\!+\! \frac{v^2}{M^2}\frac{F_\ell(u,v)}{F_0(u,v)}
\Big],
\\
F_\ell(u,v)&=&\cosh[(\frac{1}{2}M\!\!-\!\ell)\log\phi].
\end{eqnarray} 
Now, after tedious but straightforward expansion in $M^{-1}$ one finds that 
\begin{eqnarray}
\frac{F_\ell(u,v)}{F_0(u,v)}&=& 
 1-\frac{2\ell \sqrt{u^2\!+\!v^2}}{M}\tanh(\sqrt{u^2\!+\!v^2})+\frac{2\ell^2(u^2\!+\!v^2)}{M^2}
+\order(M^{-3}).
\end{eqnarray}
Hence
\begin{eqnarray}
\Xi_\ell(u,v)&=&
1
-
\frac{2\ell v^2}{M}\frac{\tanh(\sqrt{u^2\!+\!v^2})}{\sqrt{u^2\!+\!v^2}}
+
\frac{2\ell^2 v^2}{M^2}+\order(M^{-3}).
\end{eqnarray}
It follows that the equations for $\tilde{\TE}=M(1\!-\!\TE)$ and $\tilde{C}=M(1\!-\!C)$ take the form
\begin{eqnarray}
\tilde{\TE}= 
2v^2\frac{\tanh(\sqrt{u^2\!+\!v^2})}{\sqrt{u^2\!+\!v^2}}
\!-\!
\frac{2v^2}{M}\!+\!\order(M^{-2}),~~~~~~\tilde{C}=2\tilde{\TE}\!-\!\frac{4v^2}{M}
\!+\!\order(M^{-2}).~~~
\label{eq:intermediate_Etilde}
\end{eqnarray}
The dynamical equations then become
 \begin{eqnarray}
\frac{\rmd}{\rmd t}m&=& 
\tilde{\TE}\beta(h\!+\! J_0m)-2m\beta^2\Gamma^2
+\order(\frac{1}{M}),
\label{eq:dyn_M2a}
\\
\frac{\rmd}{\rmd t}\tilde{\TE}&=&  4M(\beta^2\Gamma^2\!-\!v^2)
-8\beta^3\Gamma^2m(J_0m\!+\!h)-4\beta^2\Gamma^2\tilde{\TE}+\order(\frac{1}{M}).~~~~~~~
\label{eq:dyn_M2b}
\end{eqnarray} 
What remains is to express $v$ in terms of $(m,\tilde{\TE})$, in leading two orders, by solving equation (\ref{eq:intermediate_Etilde}) for $\tilde{\TE}$ alongside our equation for $m$. 
The latter is 
\begin{eqnarray}
m
&=& 
 \frac{u\tanh(\sqrt{u^2\!+\!v^2})}{\sqrt{u^2\!+\!v^2}}+\order(M^{-2}).
 \end{eqnarray}
Equation (\ref{eq:intermediate_Etilde}) shows that $v=0$ corresponds to $\tilde{\TE}=0$, and that $\tilde{\TE}$ increases with $v^2$. 
On intermediate  timescales $\tau=M^{-1}$  we have 
 \begin{eqnarray}
\frac{\rmd}{\rmd t}m= 
\order(\frac{1}{M}),~~~~~~
\frac{\rmd}{\rmd t}\tilde{\TE}&=&  4(\beta^2\Gamma^2\!-\!v^2)
+\order(\frac{1}{M}).
\end{eqnarray} 
 Here $m$ remains constant, and $\tilde{\TE}$ evolves towards the value for which $v=\beta\Gamma+\order(M^{-1})$ (which is also the equilibrium value for $v$). Thus, in the dynamical equations (\ref{eq:dyn_M2a},\ref{eq:dyn_M2b}) describing the process on timescales with $\tau=M^{-2}$ we must substitute $v^2=\beta^2\Gamma^2+\order(M^{-1})$. So during the slow process where $m$ evolves we have always
 \begin{eqnarray}
\tilde{\TE}&=& 
2\beta^2\Gamma^2 m/u.
\end{eqnarray}
Upon insertion into (\ref{eq:dyn_M2a}) this results in a closed dynamical equation for $m$ only:
  \begin{eqnarray}
\frac{\rmd}{\rmd t}m&=& 2\beta^2\Gamma^2\Big(
\frac{\beta(h\!+\! J_0m)\tanh(\sqrt{{u}^2\!+\!\beta^2\Gamma^2})}{\sqrt{{u}^2\!+\!\beta^2\Gamma^2}}-m\Big),
\label{eq:slow_flow1}
\end{eqnarray}
without requiring approximations, and 
with $u$ to be solved from\footnote{For certain values of $m$ and $\beta\Gamma$ equation (\ref{eq:slow_flow2}) may have more than one solution $u$. In such cases the physical solution is the one with the largest absolute value.}
\begin{eqnarray}
m
&=& 
 \frac{{u}\tanh(\sqrt{{u}^2\!+\!\beta^2\Gamma^2})}{\sqrt{{u}^2\!+\!\beta^2\Gamma^2}}.
 \label{eq:slow_flow2} \end{eqnarray}
 In equilibrium we recover from (\ref{eq:slow_flow1},\ref{eq:slow_flow2}) the correct equilibrium state
 (\ref{eq:thermodynamic_state}), with $u=\beta(J_0m\!+\!h)$. Comparison with Equation (10) in \cite{Inoue1} reveals, apart from a harmless difference in time units, that the approximation of \cite{Inoue1} (used also in \cite{Inoue2,Bapst,Arai}) implies replacing ${u}$ {\em at any time} by $\beta(J_0m\!+\!h)$. 
 While this indeed holds in equilibrium, the approximation may be dangerous far from equilibrium. 
 
 \begin{figure}[t]
\unitlength=0.3mm
\hspace*{-5mm}
\begin{picture}(400,175)
\put(0,0){\includegraphics[width=250\unitlength]{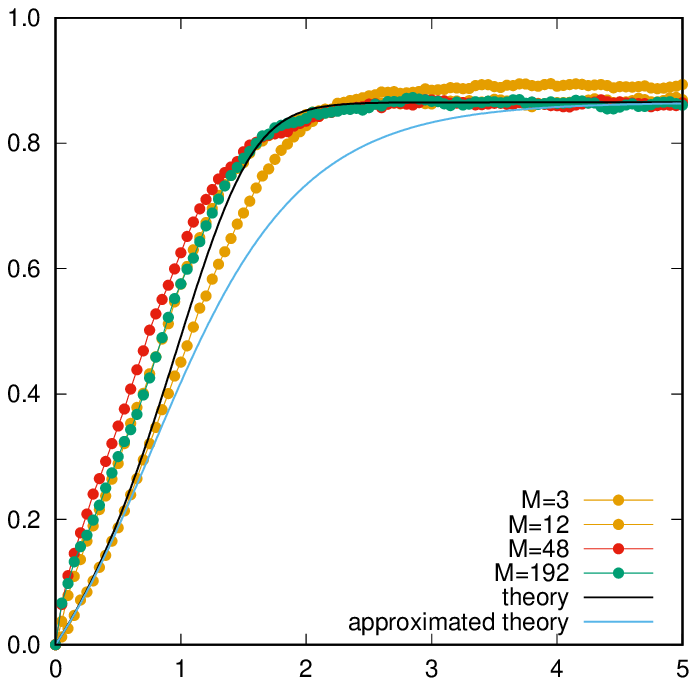}}
\put(200,0){\includegraphics[width=250\unitlength]{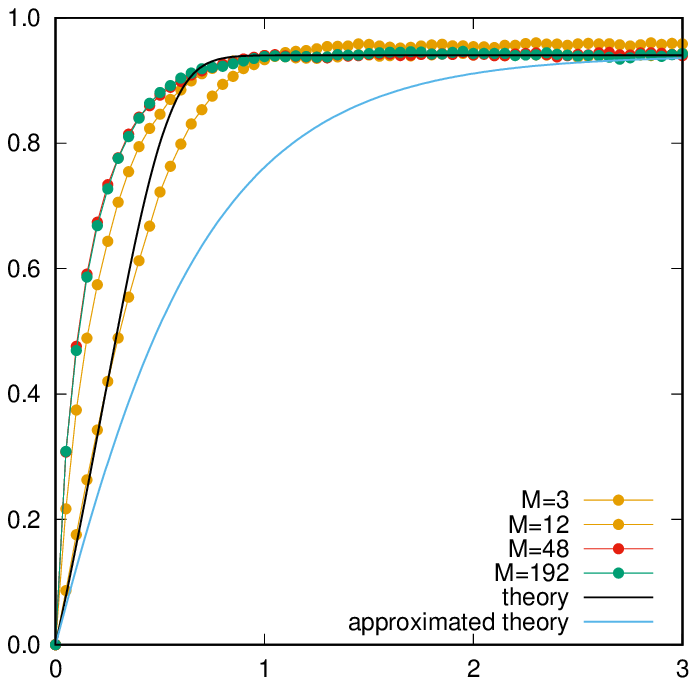}}
\put(117,-7){$t$}\put(317,-7){$t$}\put(15,90){$m$}

\put(160,70){$h\!=\!0.1$}\put(360,70){$h\!=\!0.5$}
\end{picture}
\vspace*{3mm}

\caption{Theory versus computer simulations of the microscopic  process (\ref{eq:qMCMC1},\ref{eq:qMCMC2}) for the Trotter representation of the system with Hamiltonian $H=-(J_0/N)\sum_{i<j}\sigma_i^z \sigma_j^z-\sum_i(h\sigma_i^z+\Gamma\sigma_i^x)$, with $N\!=\!10000$ and $M\!\in\!\{3,12,48,192\}$.  In all cases $J_0\!=\!1$, $T\!=\!\Gamma\!=\!0.5$, and $\tau\!=\!1/M^2$ (so time units correspond to $NM^3$ attempted moves per spin). Left figure: magnetization versus time for $h\!=\!0.1$; right figure: the same for $h\!=\!0.5$. The simulation data are shown as connected markers. The black curve is the theoretical prediction, i.e. the solution of (\ref{eq:slow_flow1},\ref{eq:slow_flow2}). The light blue curve is the approximated theory of \cite{Inoue1}, obtained by solving (\ref{eq:slow_flow1}) with the equilibrium value $u\!=\!\beta(J_0m+h)$.}

 \label{fig:simulations}
\end{figure} 

 In Figure \ref{fig:simulations} we test the predictions of (\ref{eq:slow_flow1},\ref{eq:slow_flow2}) against numerical simulations of the process (\ref{eq:qMCMC1},\ref{eq:qMCMC2}). The approximate co-location of the simulation curves for widely varying values of $M$ confirms that $\tau=\order(1/M^2)$ (inferred from the the dynamical theory) indeed captures the characteristic timescale of the macroscopic process. 
 Second, while not showing perfect agreement with the simulation data, which is not expected in view of the probability equipartitioning assumption used to close the macroscopic dynamical equations, away from stationarity the full theory (\ref{eq:slow_flow1},\ref{eq:slow_flow2}) is reasonably accurate, and  improves upon the approximation proposed in \cite{Inoue1}. 

\section{Discussion}

In this chapter we aimed to explain the basic ideas and assumptions behind the DRT strategy for deriving and closing  macroscopic dynamical equations, and its application to the types of spin systems used in quantum annealing with transverse fields. We have focused on technicalities relating to commutation of the limits $N\to\infty$ and $M\to\infty$, the possible choices of macroscopic observables, the distinct $M$-dependent timescales in the evolution of the Trotter system, and on how an additional approximation made in earlier studies can be avoided, leading to a more precise dynamical theory.  
We have tested the theoretical predictions of the theory against numerical MCMC simulations of a ferromagnetic quantum system \cite{Chayes} with  transverse external fields in Trotter representation, and found good agreement. 

In the examples used in this text there was no disorder, so we could work with the dynamical laws (\ref{eq:closed_macroscopic_flow}). If, in contrast, there is disorder in the problem, 
the macroscopic laws are to be averaged over its realization, and the main tool will be (\ref{eq:closed_averaged_macroscopic_flow}).  For models with random interactions, doing this disorder average is however relatively painless, and will not make the dynamical theory significantly more complicated. 

We hope that this introduction to the method may aid the development of further analytical studies of the macroscopic dynamics of quantum annealing, including more realistic quantum systems with disordered spin interactions or with interactions on finitely connected graphs, and more precise descriptions in which the macroscopic dynamical observables are functions \cite{DRT3,Mozeika1,Mozeika2} instead of scalars.

\begin{acknowledgement}
The authors are very grateful for stimulating discussions with Professors Hidetoshi Nishimori and  Kazuyuki Tanaka, and with Mr Shunta Arai.
\end{acknowledgement}

%\begin{figure}[t]
%\sidecaption
%\includegraphics[scale=.65]{figure}
%\caption{If the width of the figure is less than 7.8 cm use the \texttt{sidecapion} command to flush the caption on the left side of the page. If the figure is positioned at the top of the page, align the sidecaption with the top of the figure -- to achieve this you simply need to use the optional argument \texttt{[t]} with the \texttt{sidecaption} command}
%\label{fig:1}      
%\end{figure}

%%%%%%%%%%%%%%%%%%%%%%%%%%%%%%%%

\appendix

\section{Mathematical identities}

Here we list some basic properties of relevant transfer matrices and expectation values in the single-site Trotter system. 
The transfer matrix and its eigenvalues are
\begin{eqnarray}
\bK&=&\left(\!\begin{array}{cc} \rme^{y+x} & \rme^{-y}\\ \rme^{-y} & \rme^{y-x}\end{array}\!\right),~~~~~~
\lambda_\pm=\rme^y\Big[\cosh(x)\pm \sqrt{\sinh^2(x)+\rme^{-4y}}\Big].
\end{eqnarray}
The corresponding normalized eigenvectors are
\begin{eqnarray}
|+\ket&=& \frac{1}{L} \Big(\rme^{-2y}, \sqrt{\sinh^2(x)+\rme^{-4y}}-\sinh(x)\Big),
\\
|-\ket&=& \frac{1}{L} \Big(\sqrt{\sinh^2(x)+\rme^{-4y}}-\sinh(x),-\rme^{-2y}\Big),
\\
L^2&=& \rme^{-4y}+ \Big(\sqrt{\sinh^2(x)+\rme^{-4y}}-\sinh(x)\Big)^2.
\end{eqnarray}
From these expressions one can find $\bra \pm|\sigma^z|\pm\ket=
\pm \sinh(x)/\sqrt{\sinh^2(x)\!+\!\rme^{-4y}}$, and compute the following observables (with $\phi=\lambda_-/\lambda_+$):
\begin{eqnarray}
\frac{\sum_{s_1\ldots s_M}s_1 \prod_{k=1}^M \!K_{s_k s_{k+1}}}{\sum_{s_1\ldots s_M}\prod_{k=1}^M \!K_{s_k s_{k+1}}}
&=&-  \frac{\sinh(x)\tanh[\frac{1}{2}M\log\phi]}{\sqrt{\sinh^2(x)\!+\!\rme^{-4y}}},
\\
\frac{\sum_{s_1\ldots s_M}s_1 s_2\prod_{k=1}^M \!K_{s_k s_{k+1}}}{\sum_{s_1\ldots s_M}\prod_{k=1}^M \!K_{s_k s_{k+1}}}
&=&
\frac{\sinh^2(x)}{\sinh^2(x)\!+\!\rme^{-4y}}
\!+\!\frac{\cosh[(\frac{1}{2}M\!\!-\!1)\log\phi]}{\cosh[\frac{1}{2}M\log \phi]}
\frac{\rme^{-4y}}{\sinh^2(x)\!+\!\rme^{-4y}}\nonumber
\\[-1mm]
&&\\
 \frac{\sum_{s_1\ldots s_M}s_1 s_3\prod_{k=1}^M \!K_{s_k s_{k+1}}}{\sum_{s_1\ldots s_M}\prod_{k=1}^M \!K_{s_k s_{k+1}}}
&=&\frac{\sinh^2(x)}{\sinh^2(x)\!+\!\rme^{-4y}}\!+\!\frac{\cosh[(\frac{1}{2}M\!\!-\!2)\log\phi]}{\cosh[\frac{1}{2}M\log \phi]}
\frac{\rme^{-4y}}{\sinh^2(x)\!+\!\rme^{-4y}}\nonumber\\[-1mm]
&&
\end{eqnarray}


\begin{thebibliography}{99}

\bibitem{Nishimori}
Kadowaki, T. and Nishimori, H.: Quantum annealing in the transverse Ising model. Phys. Rev. E 58 (1998) 5355-5363.

\bibitem{InoueReview}
Inoue, J.I.: Infinite-range transverse field Ising models and quantum computation. Eur. Phys. J. Special Topics 224 (2015) 149-161.  

\bibitem{InoueBook}
Suzuki, S., Inoue, J.I. and Chakrabarti, B.K.: Quantum Ising Phases and transitions in Transverse Ising Models. Springer Lecture Notes in Physics 862, 2nd Ed. (2013). 

\bibitem{Suzuki} 
Suzuki, M.: Relationship between $d$-dimensional quantal spin systems and $(d+1)$-dimensional Ising systems. Prog. Theor. Phys. 56 (1976) 1454-1469.

\bibitem{Bedeaux}
Bedeaux, D., Lakatos-Lindenberg, K. and Shuler, K.E.: On the relation between Master equations and random walks and their solutions. J. Math. Phys. 12 (1971)  2116-2123.

\bibitem{Ohzeki}
Ohzeki, M.: Quantum Monte Carlo simulation of a particular class of non-stoquastic Hamiltonians in quantum annealing. Sci. Rep. 7 (2017) 41186.

\bibitem{Inoue1}
Inoue, J.I.: Deterministic flows of order parameters in the stochastic processes of quantum Monte Carlo method. J. Phys. Conf. Ser. 233 (2010) 012020.

\bibitem{Inoue2}
Inoue, J.I.: Pattern-recalling processes in quantum Hopfield networks far from saturation. J. Phys. Conf. Ser. 297 (2011) 012012.

\bibitem{Bapst}
Bapst, V. and Semerjian, G.: Thermal, quantum and simulated quantum annealing: analytical comparisons for simple models.  J. Phys. Conf. Ser. 473 (2013)  012011. 

\bibitem{Arai}
Arai, S., Ohzeki, M. and Tanaka, K.: Dynamics of order parameters in nonstoquastic Hamiltonians in the adaptive quantum Monte Carlo method. Phys. Rev. E 99 (2019) 032120.

\bibitem{Chayes}
Chayes, L., Crawford, N., Ioffe, D. and Levit, A.: The phase diagram of the quantum Curie-
Weiss model. J. Stat. Phys. 133 (2008) 131-149. 

\bibitem{DRT1}
Coolen, A.C.C. and Sherrington, D: Dynamics of fully connected attractor neural networks near
saturation.  Phys. Rev. Lett. 71 (1993) 3886-3889.

\bibitem{DRT2}
Coolen, A.C.C. and Sherrington, D.:  Order parameter flow in the SK spin-glass I: replica
symmetry. J. Phys. A 27 (1994) 7687-7707.

\bibitem{DRT3}
Laughton, S.N., Coolen, A.C.C. and Sherrington, D.: Order-parameter flow in the SK spin-glass II: inclusion of
microscopic memory effects. J. Phys. A 29 (1996)
763-786.

\bibitem{Nishimori2}
Nishimori, H. and Nonomura, Y.: Quantum effects in neural networks. J. Phys. Soc. Jpn. 65 (1996) 3780-3796.

\bibitem{replicas1}
M\'{e}zard, M., Parisi, G. and Virasoro, M.A.: Spin glass theory and beyond. Singapore: World Scientific (1987).

\bibitem{replicas2}
Nishimori, H.: Statistical physics of spin glasses and information processing. Oxford University Press (2001).

\bibitem{RFIC}
Bruinsma, R. and Aeppli, G.: One-dimensional Ising model in a random field. Phys. Rev. Lett. 50 (1983) 1494-1497.

\bibitem{Mozeika1}
Mozeika, A. and Coolen, A.C.C.: Dynamical replica analysis of processes on finitely connected random graphs:
I. Vertex covering. J. Phys. A 41 (2008) 115003.

\bibitem{Mozeika2}
Mozeika, A. and Coolen, A.C.C.: Dynamical replica analysis of processes on finitely connected random graphs:
II. Dynamics in the Griffiths phase of the diluted Ising ferromagnet. J. Phys. A 42 (2009) 195006.

\end{thebibliography}
\end{document}